\documentclass[11pt]{article}

\usepackage[latin1]{inputenc}
\usepackage[T1]{fontenc}
\usepackage[english]{babel}
\usepackage{amsmath,amssymb,amsthm}
\usepackage{mathrsfs} 
\usepackage{graphicx}
\usepackage{bbm}
\usepackage{color}
\usepackage{comment}
\usepackage[textsize=small]{todonotes}       
\usepackage{color}
\usepackage{enumitem}

\usepackage{hyperref}
\newtheorem{theorem}{Theorem}[section]

\newtheorem{proposition}[theorem]{Proposition}

\newtheorem{remark}{Remark}[section]

\newcommand{\be}{\begin{equation}}
\newcommand{\ee}{\end{equation}}

\numberwithin{equation}{section}

\newcommand{\rem}[1]{}

\def\1{{\mathchoice {1\mskip-4mu\mathrm l}      
{1\mskip-4mu\mathrm l}
{1\mskip-4.5mu\mathrm l} {1\mskip-5mu\mathrm l}}}

\begin{document}

\title{
Large spin large deviations for interacting particle systems}
\author{
Cristian Giardin\`a$^{\textup{{\tiny(a)}}}$,
Tomohiro Sasamoto$^{\textup{{\tiny(b)}}}$
\\
{\small $^{\textup{(a)}}$
Dipartimento FIM, Universit\`a di Modena e Reggio Emilia},
{\small via Campi 213/A, 41125 Modena}
\\
{\small $^{\textup{(b)}}$
Department of Physics, Institute of Science Tokyo},
{\small 2--12--1 Ookayama, Meguro-ku, Tokyo 152-8550 }
}

\maketitle


\pagenumbering{arabic}


\begin{abstract}
\noindent
We propose a novel framework for analyzing the sample-path large deviations of interacting particle systems directly on a lattice by considering a large-spin limit ($s \to \infty$). Through an appropriate scaling of variables, the process concentrates around its deterministic trajectory, governed by the lattice diffusion equation. 
We quantify the exponential cost of observing atypical trajectories by explicitly computing the path-space large deviation rate function from the scaled microscopic generator, based on a theory due to Feng and Kurtz. The rate function is described in terms of a Hamiltonian on a lattice in terms of classical spin variables.  Notably, for our main model, the symmetric harmonic process, the corresponding  Hamiltonian is identified with the classical integrable non-compact Ishimori spin chain.

Furthermore, we show that our large spin large deviation theory is fully consistent with traditional Macroscopic Fluctuation Theory (MFT), smoothly recovering the standard MFT Hamiltonian in the spatial continuous limit. We also perform a Poisson-noise Martin--Siggia--Rose (MSR) computation to derive the non-linear lattice Hamiltonian directly from microscopic dynamics, use a relative entropy argument to obtain the density-current action, and  establish the lattice fluctuation theorem alongside its symmetry origin. Overall, our framework provides a naturally regularized, lattice analogue of MFT that circumvents spatial ultraviolet singularities and possible continuous density blow-ups.


We also apply our new scheme to partial symmetric exclusion process, symmetric inclusion processes, Brownian energy process and integrable heat conduction model,  showing how their large-deviation Hamiltonians relate to classical compact and non-compact spin chains.




\end{abstract}


\section{Introduction and results}
\label{intro}

%
%
%


\subsection{Background and motivation}
The macroscopic behavior of non-equilibrium interacting particle systems has long been a central theme in statistical mechanics. A groundbreaking milestone in this field was the development of Macroscopic Fluctuation Theory (MFT) by Bertini et al. \cite{BerDeSGabJonLan15}, which provides a comprehensive variational framework to characterize the large deviations of diffuse systems in non-equilibrium steady states (NESS). Conventionally, MFT operates in the continuum limit, where space and time are rescaled simultaneously, and the underlying microscopic dynamics converge to deterministic hydrodynamic equations. While MFT has achieved remarkable success in reproducing the fluctuating hydrodynamics of systems such as the Symmetric Exclusion Process (SEP), several profound mathematical and conceptual challenges remain unresolved when dealing with more complex particle systems. 

A major bottleneck arises from the continuous space limit itself. In the standard Kipnis-Olla-Varadhan (KOV) approach \cite{KipOllVar89}, the derivation of the hydrodynamic limit and the corresponding large deviation principle (LDP) requires the so-called super-exponential estimate. For systems exhibiting unbounded local state space and reversible measure with exponential tails -- such as the symmetric inclusion process -- the super-exponential estimate could not be proved. Consequently, a rigorous proof of the hydrodynamic LDP for the inclusion process and related models has remained an outstanding open problem in interacting particle systems.

To circumvent the absence of a super-exponential estimate, this paper introduces a novel alternative approach: the large-spin large deviation principle. Instead of taking the spatial continuum limit under diffusive rescaling, we fix the underlying lattice structure (such as $\mathbb{Z}$ or a finite graph) and consider models having a spin parameter $s$. By scaling the variables appropriately by $s$ and taking the large-spin limit ($s \to \infty$), the stochastic particle process concentrates around its deterministic trajectory on the lattice. We show that the cost of observing atypical trajectories is quantified by a well-defined lattice-level large deviation function, which behaves as a naturally regularized analogue of the traditional MFT.

For models not having a spin parameter $s$, one can define the scaled process introducing an ad-hoc scaling parameter $n$ such that the scaled process has a deterministic behavior in the limit $n\to\infty$. A simple example is provided in \ref{rw},
where the case of a system of independent random walkers is discussed.
See also \cite{maes2007and, pardoux2017large, agazzi2018large, monthus2019microscopic, patterson2019large,gabrielli2020dynamical} for previous work on large deviations of interacting particle systems around their fluid limit.

In this paper, we discuss two aspects of our large spin framework. 
One is to explain the framework itself and show how it works for a few models. 
The other is to discuss its foundation and connections to the traditional hydrodynamic MFT. 
In both parts, we discuss our framework by mainly focusing on the harmonic process \cite{frassek2020non}, which we introduce next.

\subsection{Harmonic process}
The harmonic process on the torus $\mathbb{T}_N$ with $N$ sites 
is a Markov process $\{\eta(t),\, t\ge 0\}$ taking values in $\mathbb{N}^{N}$, and its generator is given by 
\be
L =  \sum_{i=1}^N  L_{i,i+1}
\label{generator}
\ee
with
\begin{eqnarray}
\label{gen-harmonic}
L_{i,i+1}f(\eta) 
& = & 
\sum_{k=1}^{\eta_i} \varphi_s(k,\eta_{i}) (f(\eta -k\delta_i+k\delta_{i+1})-f(\eta)) \nonumber \\ 
&+&  
\sum_{k=1}^{\eta_{i+1}} \varphi_s(k,\eta_{i+1}) (f(\eta +k \delta_i -k\delta_{i+1})-f(\eta))
\end{eqnarray}
where the rates $\varphi_s(k,n)$ to move $k$ particles out of $n$ available ones are given by
\be
\varphi_s(k,n) = \frac{1}{k} \frac{\Gamma(n+1)}{\Gamma(n+2s)} \frac{\Gamma(n-k+2s)}{\Gamma(n-k+1)}
\ee
and $\pm k\delta_i$ denotes addition/subtraction of $k$ particles at site $i\in\mathbb{T}_N$.
In section \ref{other}, application of our scheme to a few other examples will also be briefly discussed.  

The main results in the first part of the paper are summarized as follows.  
Let us consider the scaled process $\{\eta^{(s)}(t)\,,t\ge 0\}$ defined by
\be
\label{scaled-process}
\eta^{(s)}(t) := \frac{\eta\big(2s t\big)}{2s}.
\ee
We will first show that this scaled version of the harmonic process converges as $s\to \infty$ to  deterministic trajectories. 
\begin{proposition}[Law Large Numbers]
\label{lln}
Assume that $\eta^s(0) \to x(0)$ as $s\to\infty$ in probability.
Then the process $\{\eta^{(s)}(t),\,t\ge 0\}$ weakly converges, in the Skorokhod topology, to the deterministic process $\{x(t),\, t\ge 0\}$ on $[0,\infty)^N$ which is initialized from $x(0)$ and which solves  
\be
\label{odee}
\dot{x}_i(t) = x_{i-1}(t) - 2x_i(t) + x_{i+1}(t),  \qquad\qquad i \in \{1,2,\ldots,N\}.
\ee
\end{proposition}
\noindent
In the usual hydrodynamics, LLN for the empirical density profile is typically the diffusion equation. 
Proposition \ref{lln} is a lattice analogue for large spin. 
The proof will be given in section \ref{pr_lln}.
In fact a few other processes considered in section \ref{other} and also the independent random walkers in  \ref{rw} shows the same deterministic behaviors for large $s$. 

Next in section \ref{FK}, we describe the large deviation 
around the lattice diffusion. We compute the probability of observing an atypical trajectory $(q(t))_{t\in [0,T]}$  
by applying the theory due to Feng-Kurtz \cite{FenKur06}.
\begin{theorem}[Large deviation function of the harmonic process]
\label{main-theorem}
The $\{\eta^{(s)}(t), \,t\ge 0\}$ process 
has a path-space large deviation function of the form
\be
\lim_{s\to \infty} \frac{1}{2s}\log \mathbb{P} \Big((\eta^{(s)}(t))_{t\in[0,T]} \approx (q(t))_{t\in [0,T]}\Big) = -I ((q(t))_{t\in [0,T]}). 
\ee 
The large deviation function is written in the form,  
\be
\label{ld-fct}
I ((q(t))_{t\in [0,T]}) = I_0(q(0)) + \int_{0}^T {\cal L} (q(t),\dot q(t)) dt,
\ee
where the part $I_0$ represents the cost associated to the initial condition, 
and the  Lagrangian  ${\cal L}(q,\dot q)$ is given by the  Legendre transform of a Hamiltonian ${\cal H} (q,p)$, 
\be
{\mathcal L} (q,\dot{q}) = \sup_{p} \Big\{p\cdot \dot{q} -  {\cal H} (q,p)\Big\},
\ee
and the Hamiltonian for the harmonic process is given by 
\be
\label{HamiHarm}
\mathcal{H}(q,p) = - \sum_{i=1}^N \log \bigg(\left[ 1 - q_i \left( e^{p_{i+1}-p_i}-1 \right)\right]\left[1- q_{i+1} \left(e^{p_i-p_{i+1}}-1 \right) \right]\bigg).
\ee
\end{theorem}
The classical Hamiltonian describing the rate function is written in terms of the 
canonical coordinates $(q,p)$ satisfying the  Poisson brackets
\be
\label{pqPoi}
\{q_i,q_j\} =0, \qquad\qquad \{p_i,p_j\} =0, \qquad \qquad \{q_i,p_j\}  = \delta_{i,j},
\ee
with $\delta_{i,j}$  the Kronecker delta. 
Now let us set 
\begin{equation}
\label{su11pq}
\mathcal{S}_i^+ = (1 + q_i) e^{p_i}, \quad \mathcal{S}_i^- = q_i e^{-p_i}, \quad \mathcal{S}_i^z = q_i + \frac{1}{2}
\end{equation}
which satisfy the classical $su(1,1)$ Lie algebra, 
\begin{equation}
    \{\mathcal{S}_i^+, \mathcal{S}_j^-\} = - 2\delta_{i,j}\mathcal{S}_i^z, \qquad \{\mathcal{S}_i^z, \mathcal{S}_j^{\pm}\} = \pm\delta_{i,j} \mathcal{S}_i^{\pm}.  
    \label{su11cl}
\end{equation}
In the representation \eqref{su11pq}
the Casimir is 
\be
(\mathcal{S}_i^z)^2 - \mathcal{S}_i^+\mathcal{S}_i^- = \left(q_i+\frac12\right)^2 -q_i(1+q_i) = \frac14.
\ee
Therefore, writing $\mathcal{S}_j^{x}=(\mathcal{S}_j^+ + \mathcal{S}_j^-)/2$ and $\mathcal{S}_j^{y}=(\mathcal{S}_j^+ - \mathcal{S}_j^-)/2i$, one checks that the $su(1,1)$ spin variables $\mathcal{S}_i=(\mathcal{S}_i^x,\mathcal{S}_i^y,\mathcal{S}_i^z)$ are constrained on the hyperboloid   
$({\cal S}_i^z)^2- ({\cal S}_i^x)^2 - ({\cal S}_i^y)^2  = \frac14
$.
\begin{proposition}[Ishimori spin chain representation of the Hamiltonian]
\label{HarmIshi}
In terms of the spin variables \eqref{su11pq}, the Hamiltonian for the harmonic process \eqref{HamiHarm} can be written as 
\be
\label{HamiIshi}
 {\cal H}_{\text{Ishi}}= - \sum_{i=1}^N \log\Big[2\Big(\mathcal{S}_i\cdot \mathcal{S}_{i+1}+\frac{1}{4}\Big)\Big]. 
\ee
\end{proposition}

This rewriting and a relation to a classical limit of a quantum spin chain Hamiltonian associated with the harmonic process will be explained in section \ref{Ishi}. 
The Hamiltonian \eqref{HamiIshi} is the $su(1,1)$ version of the Ishimori spin chain \cite{Hal82,Ish82}, which is known to be classically integrable.
If one takes the continuous space limit, the Ishimori Hamiltonian \eqref{HamiIshi} becomes the non-compact analogue of the Landau-Lifshitz model (LLE) \cite{FadTak07}, which is related to the MFT equation for SEP. 
The Hamilton equation for the Hamiltonian  (\ref{HamiHarm}) is a lattice analogue of the MFT equation. 
In addition the Ishimori chain is known to be a integrable discretization of LLE. 
In a similar manner to \cite{MalMorSas22}, one may be able to solve the Hamilton equation for the Hamiltonian  (\ref{HamiHarm}) and determine the rate function. 
Indeed, in a forthcoming companion paper \cite{GMSS2026p}, we will solve these equations of motion explicitly by utilizing the complete integrability of the underlying Ishimori chain. By implementing the inverse scattering transform and evaluating the corresponding action functional, we will derive the exact closed-form expression for the cumulant generating function of the integrated current, thereby providing the first exact solution to a non-compact macroscopic fluctuation framework on a fixed lattice.

We emphasize that the connection to a classical integrable system is a particular property of the harmonic process, and does not apply to other models 
such as the partial exclusion process and the inclusion process in section \ref{other}, whose Hamiltonians in $p,q$ variables read
($-$ for exclusion and $+$ for inclusion) 
\be
\label{HamiExI}
 {\cal H}(q,p) =  \sum_{i=1}^N \Big(
q_i(1\mp q_{i+1}) (e^{-(p_{i}-p_{i+1})}  - 1)  +   
(1\mp q_{i})q_{i+1} (e^{-(p_{i+1} -p_{i}) } - 1) \Big). 
\ee
Their corresponding spin chain Hamiltonians are classical lattice XXX spin chains, which are supposed to be non-integrable.  

\subsection{Discrete MFT}
In the second part of this paper,
we explore the theoretical foundations of our framework and its connections to macroscopic fluctuation phenomena.

We then establish how our lattice formulation bridges microscopic dynamics to Macroscopic Fluctuation Theory (MFT). We show that taking the continuous space limit recovers the standard continuum MFT action, confirming that our discrete formulation provides a naturally regularized counterpart to continuum large deviation theories. Then, using a lattice-level Martin--Siggia--Rose (MSR) path-integral framework with Poisson jump noise, we derive the explicit non-linear lattice Hamiltonian and its corresponding density-current action directly from microscopic conservation laws. Within this lattice setting, we prove the fluctuation theorem and analyze its microscopic symmetry origin. 


The following scheme summarizes the relation between large spin large deviation and MFT: 
\begin{equation*}
\begin{array}{r@{\quad}c@{\quad}c@{\quad}c} 
 \textbf{Lattice:} & \text{Large $s$ LD} & \longleftrightarrow & \text{Classical spin chain } \\ 
& \downarrow & & \downarrow \\ 
\textbf{Continuum:} & \text{Diffusive MFT} & \longleftrightarrow &  \text{Continuous classical spin} 
\end{array}
\end{equation*}

This scheme visually captures the primary conceptual contribution of our work. By establishing the horizontal links on the lattice level, we demonstrate that the large-$s$ scaling serves as a universal mechanism that translates stochastic particle fluctuations into classical spin field trajectories before any spatial continuum limit is taken. The vertical arrows emphasize that our regularized lattice-MFT framework successfully interpolates between discrete spin systems and the standard continuous MFT. 
For integrable models such as the harmonic process, we can add another column of Ablowitz-Ladik model and the AKNS system (or NLS), which represent further related integrable models suitable for finding exact solutions.

\subsection{Organization of the paper}
 The remainder of this paper is organized as follows. In Section~\ref{pr_lln}, we establish the law of large numbers for the large-spin scaling of the harmonic process and give a rigorous proof based on the Trotter--Kurtz theorem. Section~\ref{FK} develops the large-deviation theory for the harmonic process using the Feng--Kurtz framework and derives the corresponding lattice Hamiltonian explicitly from the scaled microscopic generator. In Section~\ref{Ishi}, we show that this Hamiltonian can be represented in terms of classical non-compact $\mathrm{su}(1,1)$ spin variables and identify it with the integrable Ishimori spin chain. We further discuss the relation between the large-deviation Hamiltonian and the classical limit of the underlying quantum spin chain. 
 

Sections \ref{cts} and \ref{mft} establish the connection between the large-spin lattice theory and conventional Macroscopic Fluctuation Theory. We first study the continuous-space limit of the lattice Hamiltonians and recover the standard MFT Hamiltonian. We then derive the MFT Hamiltonian directly from the fluctuating hydrodynamics generator using the Feng--Kurtz construction, and provide a lattice-regularized Martin--Siggia--Rose formulation based on the underlying Poisson jump processes. We also derive a density-current representation of the large-deviation action, whose continuum limit gives the conventional Gaussian MFT action. Furthermore, in section \ref{sec:lattice-ft-einstein}, we discuss the fluctuation theorem and Einstein relation at the lattice level. Section~\ref{other} extends the large-spin large-deviation framework to other interacting particle systems, including the partial symmetric exclusion process, the symmetric inclusion process, the integrable heat conduction model, and the Brownian energy process, and discusses their relations to classical compact and non-compact XXX and Ishimori spin chains. Finally, ~\ref{rw} illustrates the general construction for the simplest example of independent random walkers. 

\paragraph{Acknowledgments.}
The work of TS has been supported by JSPS KAKENHI Grants No. JP21H04432, No. JP23K22414.
CG acknowledges  support from Istituto Nazionale di Alta Matematica (INdAM) Grant No. CUP E53C23001740001. 
The authors thank T. Bodineau, D. Gabrielli, K. Mallick, F. Redig, H. Suda for useful discussions and suggestions. 
They are grateful to the Isaac Newton Institute for Mathematical Sciences for its hospitality during the program "New statistical physics in living matter: non-equilibrium states under adaptive control" (SPL), where this work was initiated. Part of this work was performed during the workshop "Rigorous Statistical Mechanics and Related Topics" at the Research Institute for Mathematical Sciences, Kyoto University, the workshops "Hydrodynamics, Fluctuations, and Noise in quantum and classical systems" (code: ICTS/hydrodynamics2025/12) and "Two and a half decades of the Macroscopic Fluctuation Theory 2026" (code: ICTS/MFTHEORY2026/09) at the International Centre for Theoretical Sciences (ICTS), and the workshop "Algebraic Methods in Probability" at the Simons Center for Geometry and Physics (SCGP). We acknowledge them for their support.

\section{The proof of law of large numbers}
\label{pr_lln}
In this section we prove Proposition \ref{lln}, the law of large numbers for the harmonic process. 

%
%
\begin{proof}
The proof is an application of the Trotter-Kurtz theorem \cite{EthKur09}. To show
convergence of the process $\{\eta^{(s)}(t),\,t\ge 0\}$ to the deterministic system $\{x(t),\, t\ge 0\}$
(where the convergence is weak convergence  in the Skorokhod topology)
it suffices to show that the generator $L^{(s)}$ of the scaled process $\{\eta^{(s)}(t)\,,t\ge 0\}$ 
converges as $s\to\infty$  to the generator 
\begin{equation}
\label{Lx}
L^X f(x) = \sum_{i=1}^N (x_{i-1} - 2x_i + x_{i+1}) \frac{\partial f}{\partial x_i}.
\end{equation}
The generator $L^{(s)}$  reads
\begin{eqnarray}
\label{harmonic-scaled-gen}
L^{(s)}f(x) =
&&
2s \sum_{i=1}^N \Big(
\sum_{k=1}^{2s x_i} \varphi_s(k,2s x_{i}) (f(x -\frac{k}{2s}\delta_i+\frac{k}{2s}\delta_{i+1})-f(x)) \nonumber \\ 
&&
\quad\quad\quad+  
\sum_{k=1}^{2s x_{i+1}} \varphi_s(k,2s x_{i+1}) (f(x +\frac{k}{2s} \delta_i -\frac{k}{2s}\delta_{i+1})-f(x))
\Big).
\end{eqnarray}
%
Taylor expanding we have
\begin{eqnarray}
L^{(s)}f(x) =
&&
2s \sum_{i=1}^N \left(
 \sum_{\ell=1}^{\infty}
  \frac{1}{\ell!} 
  \left\{\sum_{k=1}^{2s x_i} \Big(-\frac{k}{2s}\Big)^\ell   \varphi_s(k,2s x_{i})   \right\}
 \left(\frac{\partial }{\partial x_i} - \frac{\partial }{\partial x_{i+1}}\right)^\ell f(x) 
\right. \nonumber\\
&&
+\qquad\qquad \left.
 \sum_{\ell=1}^{\infty}
  \frac{1}{\ell!} 
  \left\{ \sum_{k=1}^{2s x_{i+1}}  \Big(\frac{k}{2s}\Big)^\ell \varphi_s(k,2s x_{i+1}) \right\}
 \left(\frac{\partial }{\partial x_i} - \frac{\partial }{\partial x_{i+1}}\right)^\ell f(x) 
\right) .
\end{eqnarray}
Using
\be
\sum_{k=1}^{2sx_i}k \varphi_s(k,2sx_i)= \frac{x_i}{2s}
\ee
the first-order derivative terms ($\ell = 1$)
in the expansion of $L^{(s)}$
recombine to yield the  drift of the deterministic system. To ensure the applicability of the Trotter-Kurtz theorem \cite{EthKur09}, we check that the remainder terms vanish uniformly for large $s$. For a smooth test function $f$ with bounded second derivatives, the remainder contribution is bounded by:
\begin{equation}
\left| \text{Remainder} \right| \le \frac{C}{s} \sum_{i=1}^N \left[ \sum_{k=1}^{2sx_i} k^2 \varphi_s(k, 2sx_i) + \sum_{k=1}^{2sx_{i+1}} k^2 \varphi_s(k, 2sx_{i+1}) \right]
\end{equation}
where $C = \frac{1}{4} \sup_{x} \| \nabla^2 f(x) \|_\infty$. The summation over $k^2$ satisfies:
\begin{equation}
\sum_{k=1}^\infty k^2 \varphi_s(k,2sx_i)= \frac{2s}{2s+1} x_i (x_i + 1).
\end{equation}
Since this summation converges to a finite local density function for any compact domain of $x$, the entire remainder block scales as ${O}(s^{-1})$ and vanishes in the large-$s$ limit.  

Therefore, the generator $L^{(s)}$ converges uniformly on compact sets to the first-order differential operator
$L^X$ in \eqref{Lx}.

\end{proof}

\section{Large deviations for the harmonic process}
\label{FK}

In this section we explain how the 
Hamiltonian (\ref{HamiHarm}) appears when we apply 
to the harmonic process 
\eqref{generator},\eqref{gen-harmonic} the Feng-Kurtz scheme \cite{FenKur06}, which we review briefly. For an actual proof of  Theorem \ref{main-theorem}, see the remark at the end of this section.

Suppose that a sequence of processes $\{\eta^{(n)}(t), \, t\ge 0\}$ has a deterministic behavior for large values of the parameter $n$  and that 
we are interested in the probability that the process $\{\eta^{(n)}(t), \, t\ge 0\}$  follows a generic trajectory $\{q(t), t\ge 0\}$
in the limit $n\to\infty$. 
This probability approaches zero exponentially fast and the rate is quantified by the large deviation function: 
\be
\lim_{n\to \infty} \frac{1}{n}\log \mathbb{P} \Big((\eta^{(n)}(t))_{t\in[0,T]} \approx (q(t))_{t\in [0,T]}\Big) = -I ((q(t))_{t\in [0,T]}). 
\ee 
The large deviation function is written in the form,  
\be
\label{ld-fct2}
I ((q(t))_{t\in [0,T]}) = I_0(q(0)) + \int_{0}^T {\cal L} (q(t),\dot q(t)) dt,
\ee
where the part $I_0$ represents the cost associated to the initial condition, 
and the  Lagrangian  ${\cal L}(q,\dot q)$ is given by the  Legendre transform of a Hamiltonian ${\cal H} (q,p)$, 
\be
\label{HamiLag}
{\mathcal H} (q,p) = \sup_{\dot{q}} \Big\{p\cdot \dot{q} -  {\cal L} (q,\dot q)\Big\}
\ee
where $(q,p)$ is a set of canonical coordinates satisfying the Poisson brackets \eqref{pqPoi}. 
The Hamiltonian ${\cal H} (q,p)$ can be calculated from the generator
as follows. 
Let us denote by $L^{(n)}$ the generator of the process $\{\eta^{(n)}(t), \, t\ge 0\}$.
Then Hamiltonian ${\cal H} (q,p)$ is given by 
\be
\label{Hami}
{\cal H}(q,p) = 
\lim_{n\to\infty} 
{\cal H}^{(n)}(q,p) =\lim_{n\to\infty} \frac{1}{n} e^{-npq} L^{(n)} e^{npq}. 
\ee
Compared to the usual large deviation arguments, which require tilting of measures, this provides a convenient and efficient way to find a variational formula of a rate function. 


Now let us apply the formula \eqref{Hami} to the harmonic process, by taking the large parameter $n$ to be the spin parameter $2s$. 
The generator of its scaled version is given in \eqref{harmonic-scaled-gen}.
We find
\begin{eqnarray}
 {\cal H}^{(2s)}(q,p)
 & = & 
\frac{1}{2s} e^{-2s p q} 2s \sum_{i=1}^N \Big(
\sum_{k=1}^{2s q_i} \varphi_s(k,2s q_{i}) (e^{2s p(q -  \frac{k}{2s} \delta_i + \frac{k}{2s} \delta_{i+1} ) } - e^{2s p q})  \nonumber \\
&&  \qquad \qquad \qquad
+   
\sum_{k=1}^{2s q_{i+1}} \varphi_s(k,2s q_{i+1}) (e^{2s p(q +  \frac{k}{2s} \delta_i - \frac{k}{2s} \delta_{i+1} ) } - e^{2s p q})  \Big). 
\end{eqnarray}
Simplifying, we get
\be
\label{eq:hamiltonian-simplified}
 {\cal H}^{(2s)}(q,p)
 =  
\sum_{i=1}^N \Big(
\sum_{k=1}^{2s q_i} \varphi_s(k,2s q_{i}) (e^{(p_{i+1}-p_i)k}  - 1)  
\, +   
\sum_{k=1}^{2s q_{i+1}} \varphi_s(k,2s q_{i+1}) (e^{-(p_{i+1} -p_{i})k } - 1)  \Big).
\ee
In the limit of large spins, for every fixed \(k\) we have the
asymptotic relation
\begin{equation}
 \varphi_s(k,2s q_i)
 =
 \frac{1}{k}
 \left(\frac{q_i}{1+q_i}\right)^k
 +o(1),
 \qquad s\longrightarrow\infty.
 \label{asymp}
\end{equation}
Since the upper limit in the sums in \eqref{eq:hamiltonian-simplified}
depends on \(s\), we must justify the interchange of the limit with
the summation over \(k\). For \(q_i>0\), with \(2sq_i\in\mathbb N\), a
direct computation from the definition of the rates gives, for
\(1\leq k<2sq_i\),
\be
\frac{\varphi_s(k+1,2sq_i)}{\varphi_s(k,2sq_i)}
 =
 \frac{k}{k+1}
 \frac{2sq_i-k}{2s(q_i+1)-k-1}
 \leq
 \frac{k}{k+1}
 \frac{2sq_i}{2s(q_i+1)-1}.
\ee
Moreover,
since
$
\varphi_s(1,2sq_i)
 =
 \frac{2sq_i}{2s(q_i+1)-1},
$
iterating the preceding inequality, we obtain
\be
 \varphi_s(k,2sq)
 \leq
 \frac{1}{k}
 \left(
   \frac{2sq}{2s(q+1)-1}
 \right)^k.
\ee
Consider a compact subset of the domain
\be
 \frac{q_i}{1+q_i}e^{p_{i+1}-p_i}<1,
 \qquad
 \frac{q_{i+1}}{1+q_{i+1}}e^{-(p_{i+1}-p_i)}<1.
\ee
Since
$
 \frac{2sq}{2s(q+1)-1}
 \longrightarrow
 \frac{q}{1+q}
$
uniformly for \(q\) in compact sets, there exist \(s_0>0\) and
\(0<r<1\) such that, for all \(s\geq s_0\),
\be
 \frac{2sq_i}{2s(q_i+1)-1}e^{\max\{p_{i+1}-p_i,0\}}\leq r,
 \qquad
 \frac{2sq_{i+1}}{2s(q_{i+1}+1)-1}e^{\max\{-(p_{i+1}-p_i),0\}}\leq r.
\ee
It follows that
\be
\varphi_s(k,2sq_i)\left|e^{k(p_{i+1}-p_i) }-1\right|
\leq
 \frac{1}{k}
 \left[
   \left(
     \frac{2sq_i}{2s(q_i+1)-1}
     e^{\max\{p_{i+1}-p_i,0\}}
   \right)^k
   +
   \left(
     \frac{2sq_i}{2s(q_i+1)-1}
   \right)^k
 \right]
 \leq
 \frac{2r^k}{k}.
\ee
The same estimate holds for the second sum involving $q_{i+1}$. Since
$\sum_{k\geq1}\frac{r^k}{k}<\infty,
$
the dominated-convergence theorem applies. Using \eqref{asymp} and the identity 
\be
-\log(1-y) = \sum_{k=1}^{\infty} \frac{y^k}{k},
\ee
we
therefore obtain
\begin{eqnarray}
\lim_{s\to\infty} \sum_{k=1}^{2sq_i}
\varphi_s(k,2sq_i)
\left(e^{k(p_{i+1}-p_i)}-1\right)
 &=&
\sum_{k=1}^{\infty}
 \frac{1}{k}
 \left(\frac{q_i}{1+q_i}\right)^k
\left(e^{k(p_{i+1}-p_i)}-1\right)
 \nonumber\\
 &=&
 -\log\left(
  1-\frac{q_i}{1+q_i}e^{p_{i+1}-p_i}
 \right)
 +
 \log\left(
  1-\frac{q_i}{1+q_i}
 \right),
\label{eq:forward-limit}
\end{eqnarray}
and
\be
\lim_{s\to\infty}
\sum_{k=1}^{2sq_{i+1}}
\varphi_s(k,2sq_{i+1})
 \left(e^{-k(p_{i+1}-p_i)}-1\right)
 \nonumber\\
 =
 -\log\left(
  1-\frac{q_{i+1}}{1+q_{i+1}}
       e^{-(p_{i+1}-p_i)}
 \right)
 +
 \log\left(
  1-\frac{q_{i+1}}{1+q_{i+1}}
 \right).
\label{eq:backward-limit}
\ee
The convergence is locally uniform on the above domain. Summing the
two contributions over all bonds gives
\begin{eqnarray}
 {\cal H}(q,p)  = \lim_{s\to\infty} {\cal H}^{(2s)}(q,p) 
& = & 
\sum_{i=1}^N \Big(
- \log\left(1-  \frac{q_i}{q_i+1} e^{p_{i+1}-p_i}  \right) + \log\left(1-  \frac{q_i}{q_i+1}  \right) \nonumber  \\
&&  \qquad     
- \log\left(1-  \frac{q_{i+1}}{q_{i+1}+1} e^{-(p_{i+1}-p_i)}  \right) + \log\left(1-  \frac{q_{i+1}}{q_{i+1}+1}  \right) \Big) 
\end{eqnarray}
Doing the algebra one arrives at \eqref{HamiHarm}.

\begin{remark}[Effective domain of the Hamiltonian]
The preceding argument proves locally uniform convergence on compact
subsets of the effective domain
\begin{equation}
 e^{p_{i+1}-p_i}<1+\frac{1}{q_i},
 \qquad
 e^{-(p_{i+1}-p_i)}<1+\frac{1}{q_{i+1}},
 \qquad i\in\mathbb T_N.
 \label{eq:converg}
\end{equation}
These conditions characterize the region in which the exponential
moments of the limiting jump currents are finite. Outside
this effective domain, the Hamiltonian is understood as an
extended-valued function taking the value \(+\infty\). The cases in
which \(q_i=0\) or \(q_{i+1}=0\) follow directly by continuity.
\end{remark}


\begin{remark}[Relation to the WKB approximation]
The Feng--Kurtz approach to large deviations is closely related
to the WKB approximation in quantum mechanics.
Consider a quantum Hamiltonian
$\widehat H=\widehat H(q,-i\hbar\partial_q)$
and the Schr\"odinger equation
$i\hbar\partial_t\psi=\widehat H\psi$.
With the WKB ansatz
\begin{equation}
  \psi(q,t)=a(q,t)\exp\left(\frac{i}{\hbar}S(q,t)\right),
\end{equation}
the leading order in $\hbar$ gives the Hamilton--Jacobi equation
\begin{equation}
  \partial_t S+H(q,\partial_q S)=0.
\end{equation}
Here $H(q,p)$ is the classical Hamiltonian obtained from
$\widehat H$ by replacing $-i\hbar\partial_q$ with $p$
and taking $\hbar\to0$.
Equivalently, it can be expressed as
\begin{equation}
  H(q,p)=
  \lim_{\hbar\to0}
  e^{-ipq/\hbar}\widehat H
  \left(e^{ipq/\hbar}\right).
\end{equation}
This expression makes the analogy with the Feng--Kurtz
formula~\eqref{Hami} explicit: in both cases, a classical
Hamiltonian is obtained from an operator through exponential
conjugation and an appropriate asymptotic limit.

Note, however, a cenceptual difference that, in quantum mechanics, the classical Hamiltonian is already
encoded in the quantum operator, whereas in the Feng--Kurtz
approach it emerges directly from the transition rates
of the stochastic process.
See \cite{jona1985stochastic} for a discussion about the relation between semiclassical limit of quantum mechanics and large deviations of stochastic processes.
\end{remark}


\begin{remark}[Mathematical status of the large-$s$ LDP convergence]
While the law of large numbers in Section \ref{lln} is rigorously established via the linear Trotter-Kurtz theorem, the full mathematical proof of the sample-path large deviation principle (LDP) via the Feng-Kurz framework requires verifying the operator convergence of the non-linear semigroup associated to the generator $H^{(2s)} f = \frac{1}{2s} ( e^{-2sf} L^{(s)} e^{2sf} )$. According to the general theory developed in \cite{FenKur06}, this requires proving that the sequence of non-linear operators converges uniformly on an appropriate test function space, and that the resulting limiting Hamiltonian operator satisfies the comparison principle (uniqueness of the viscosity solutions to the associated Hamilton-Jacobi equations), which we have not established yet. 
However, for our harmonic process on a finite graph, we can prove the path space LDP by a different approach, adapting the traditional approach \cite{KipOllVar89,KipnisLandim1999}, see \cite{GS2026p}. 
\end{remark}


\section{Connection to the Ishimori spin chain}
\label{Ishi}
In this section we first prove Proposition \ref{HarmIshi}, i.e. show that the Hamiltonian \eqref{HamiHarm} in terms of $p,q$ variables for the Harmonic process can be written as the Ishimori spin chain Hamiltonian \eqref{HarmIshi} in terms of the $su(1,1)$ spins (\ref{su11pq}), and then discuss its relation to classical limit of quantum spin chains. 

First we observe that, by expanding the product inside the logarithm, the Hamiltonian \eqref{HamiHarm}
can be written as
\be
\label{HamiHarmold}
\mathcal{H}(q,p) = - \sum_{i=1}^N \log \left[ 1 + q_i(1 + q_{i+1}) \left( 1 - e^{p_{i+1}-p_i} \right) + q_{i+1}(1 + q_i) \left( 1 - e^{p_i-p_{i+1}} \right) \right].
\ee
Then, the proof of Proposition \ref{HarmIshi} is simply due to a rewriting of the spin scalar product in terms of $p,q$ variables using the transformation \eqref{su11pq} as 
\begin{eqnarray}
2 \left({\cal S}_i \cdot {\cal S}_{i+1}+\frac14\right) 
&= & 
2 \left({\cal S}_i^z {\cal S}_{i+1}^z - \frac12{\cal S}_i^+{\cal S}_{i+1}^- - \frac12{\cal S}_i^-{\cal S}_{i+1}^++\frac14\right)\nonumber\\
& = & 
2\left((q_i+\frac12)(q_{i+1}+\frac12) - \frac12(1+q_i)q_{i+1} e^{p_i-p_{i+1}} - \frac12 q_{i}(1+q_{i+1}) e^{p_{i+1}-p_i}+\frac14\right) \nonumber\\
&=&
q_i(1 + q_{i+1}) \left( 1 - e^{p_{i+1}-p_i} \right) + q_{i+1}(1 + q_i) \left( 1 - e^{p_i-p_{i+1}} \right) 
\end{eqnarray}

The Ishimori spin chain with Hamiltonian \eqref{HamiIshi} is  classically integrable \cite{Ish82}. 
Moreover this classical Hamiltonian is known to appear as a classical limit of the integrable quantum spin chain  $\text{XXX}_s$ Hamiltonian  \cite{Hal82}. The quantum $\text{XXX}_s$ integrable Hamiltonian with $\mathfrak{su}(1,1)$ spins can be written in the form  
\be
H = -\sum_{i=1}^N 2(\psi(\mathbb{S}_{i,i+1}) -\psi(2s)),
\ee
where the operator $\mathbb{S}_{i,i+1}$ is defined  via 
\be
\mathbb{S}_{i,i+1}(\mathbb{S}_{i,i+1}-1) = 2s(s-1) + 2 S_{i}\cdot S_{i+1}
\ee 
with
\be
S_{i}\cdot S_{i+1} = S_i^z S_{i+1}^z - \frac12 (S_i^+ S_{i+1}^- + S_i^- S_{i+1}^+).
\ee
Here $S_i^{\pm}, S_i^z$ satisfy the commutator relations of the $\mathfrak{su}(1,1)$ algebra,  
\begin{equation}
    [S_i^+, S_j^-] = -2\delta_{i,j}S_i^z, \qquad [S_i^z, S_j^{\pm}] = \pm\delta_{i,j} S_i^{\pm}.  
    \label{su11qu}
\end{equation}
Moreover, the harmonic process arises from the 
integrable $\text{XXX}_s$ spin chain in a (stochastic) representation of the $\mathfrak{su}(1,1)$ Lie algebra \cite{frassek2020non}.  
Thus the fact that the LD Hamiltonian is the classical limit of the quantum spin chain Hamiltonian for the Markov generator can be interpreted as a manifestation of this.

As already mentioned in introduction, the continuous space limit of the Ishimori Hamiltonian \eqref{HamiIshi} becomes the Landau-Lifshitz model (LLE) \cite{FadTak07}, which is related to the MFT equation for SEP. The exact solution for the latter was achieved in \cite{MalMorSas22} by a mapping to the AKNS system, which is a two variable version of the non-linear Schr\"odinger equation (NLS). As a discrete analogue of this, the exact solution for the large deviation of the harmonic process can be obtained by a mapping  to the Ablowitz-Ladik model, which is a discrete version of the NLS \cite{GMSS2026p}.

\section{Continuous space limit}
\label{cts}

Our large-spin large deviation theory is formulated on a lattice.
In this section, we show that the continuous space limit of the
lattice Hamiltonian yields the standard MFT Hamiltonian.
We also recover the latter by formally applying the Feng--Kurtz
scheme to the fluctuating hydrodynamic equation.

\subsection{Continuous space limit of Hamiltonians}
\label{subsec:continuum-hamiltonians}

We now show that the lattice Hamiltonian obtained from
the large-spin large deviation principle reduces to the
standard MFT Hamiltonian in the continuous space limit.

Consider a periodic lattice with $N$ sites and introduce
smooth continuum fields $\rho(u)$ and $H(u)$ by setting
    $q_i=\rho(i/N),
    p_i=H(i/N)$,
where $u\in\mathbb T=\mathbb R/\mathbb Z$.
Under the diffusive time scaling,
the continuum Hamiltonian
is obtained as
\begin{align}
    \mathcal H_{\mathrm{MFT}}[\rho,H]
    =
    \lim_{N\to\infty}N\mathcal H_N(q,p).
    \label{eq:continuum-H-scaling}
\end{align}

For the harmonic process, the Hamiltonian
\eqref{HamiHarm} is a sum of the local term,
\begin{align}
    \mathcal H_{i,i+1}
    =
    -\log\left[
        1-q_i(e^{p_{i+1}-p_i}-1)
    \right]
    -\log\left[
        1-q_{i+1}(e^{p_i-p_{i+1}}-1)
    \right].
    \label{eq:continuum-harmonic-bond}
\end{align}
Expanding in the momentum difference
$p_{i+1}-p_i$, we obtain
\begin{align}
    \mathcal H_{i,i+1}
    &=
    (q_i-q_{i+1})(p_{i+1}-p_i)
    +\frac12
    \left[
        q_i(1+q_i)+q_{i+1}(1+q_{i+1})
    \right](p_{i+1}-p_i)^2
    +O((p_{i+1}-p_i)^3).
    \label{eq:continuum-harmonic-expansion}
\end{align}
For smooth fields, both $q_{i+1}-q_i$ and
$p_{i+1}-p_i$ are of order $N^{-1}$.
Substituting their leading continuum approximations
into \eqref{eq:continuum-harmonic-expansion},
we find
\begin{align}
    \lim_{N\to\infty}N\mathcal H_N(q,p)
    &=
    \int_{\mathbb T}du\,
    \left[
        -\partial_u\rho\,\partial_uH
        +\rho(1+\rho)(\partial_uH)^2
    \right].
    \label{eq:continuum-harmonic-final}
\end{align}
This agrees with the standard MFT Hamiltonian
\begin{align}
    \mathcal H_{\mathrm{MFT}}[\rho,H]
    =
    \int_{\mathbb T}du\,
    \left[
        -D(\rho)\partial_u\rho\,\partial_uH
        +\frac12\sigma(\rho)(\partial_uH)^2
    \right],
    \label{eq:continuum-MFT-general}
\end{align}
with
\begin{align}
    D(\rho)=1,~
    \sigma(\rho)=2\rho(1+\rho).
    \label{eq:continuum-harmonic-transport}
\end{align}

The same calculation applies to the partial symmetric
exclusion and symmetric inclusion processes with
Hamiltonians \eqref{HamiExI}, yielding $D(\rho)=1$ and
$\sigma(\rho)=2\rho(1-\rho)$ or $2\rho(1+\rho)$,
respectively.
Note that the two quantities, $D$ and $\sigma$, and hence the MFT equations of the harmonic process and the inclusion process are the same. 
At the MFT level, these two models are regarded as the same system but one can study the difference of the two models, for example the effects of integrability still at the level of large $s$ large deviation. 

The continuous space limit can also be taken directly
in the classical spin-chain representation.
For the harmonic process, the Ishimori Hamiltonian \eqref{HamiIshi}
reduces to the non-compact Landau--Lifshitz Hamiltonian
\begin{align}
    \mathcal H_{\mathrm{LL}}
    =
    \int_{\mathbb T}du\,
    \partial_u S\cdot
    \partial_uS,
    \label{eq:continuum-LL}
\end{align}
where the scalar product is defined with the
non-compact spin metric. This expression is equivalent
to \eqref{eq:continuum-harmonic-final} under the
spin parametrization \eqref{su11pq}.
The corresponding integrable structures are related
to the Ablowitz--Ladik model and AKNS equation,
respectively.

\subsection{Derivation of the MFT Hamiltonian via the Feng--Kurtz scheme}
\label{subsec:mft-fk}

We now show that the standard MFT Hamiltonian can also
be obtained by formally applying the Feng--Kurtz scheme
to the fluctuating hydrodynamic equation
\begin{align}
   \label{eq:mft-spde}
    \partial_t \rho(x,t)
    =    \partial_x[D(\rho)\partial_x \rho]
    -\frac{1}{\sqrt{N}}   \partial_x[\sqrt{\sigma(\rho)}\,\xi(x,t)],
\end{align}
where $\xi(x,t)$ is space-time Gaussian white noise.
The corresponding formal Markov generator acts on a
functional $F[\rho]$ as
\begin{align}
    L_N F[\rho]
    &=
    \int_0^1 dx\,
\partial_x[D(\rho)\partial_x \rho]\,
    \frac{\delta F}{\delta \rho(x)}
    +
    \frac{1}{2N}
    \int_0^1 dx\int_0^1 dy\,
    K_\rho(x,y)
    \frac{\delta^2 F}{\delta \rho(x)\delta \rho(y)},
    \label{eq:mft-generator}
\end{align}
where
\begin{align}
    K_\rho(x,y)
    =
    -\partial_x\left[       \sigma(\rho(x))\partial_x\delta(x-y)
    \right].
    \label{eq:mft-covariance}
\end{align}
We assume periodic boundary conditions throughout.

Note that functional derivatives of 
\begin{align}
    F_H[\rho]
    =
    \exp\left[N\int_0^1 dx\,H(x)\rho(x)\right].
    \label{eq:mft-exponential}
\end{align}
are given by
\begin{align}
    \frac{\delta F_H}{\delta \rho(x)}
    =NH(x)F_H,
    \qquad
    \frac{\delta^2F_\rho}{\delta \rho(x)\delta \rho(y)}
    =N^2H(x)H(y)F_H.
    \label{eq:mft-functional-derivatives}
\end{align}
Applying the Feng--Kurtz formula formally to the generator \eqref{eq:mft-generator} and 
substituting these expressions, we find
\begin{align}
    \lim_{N\to\infty}
    \frac{1}{N}F_H[\rho]^{-1}L_NF_H[\rho]
    &=
    \int_0^1 dx\,
H(x)\partial_x[D(\rho)\partial_x \rho]
    +
    \frac12
    \int_0^1 dx\int_0^1 dy\,
    p(x)K_\rho(x,y)H(y)
    \nonumber\\
    &=
    \int_0^1 dx\,
    \left[
        -D(\rho)\partial_x\rho\,\partial_xH
        +\frac12\sigma(\rho)(\partial_xH)^2
    \right]
    =
    \mathcal H_{\mathrm{MFT}}[\rho,H].
    \label{eq:mft-fk-hamiltonian}
\end{align}
In the last step, we used integration by parts
and the definition \eqref{eq:mft-covariance}.

This calculation is formal because the continuum SPDE    \eqref{eq:mft-spde}
involves multiplicative space-time white noise, and its well-definedness is not at all trivial and is still being under investigation \cite{dirr2020conservative}.
Nevertheless, it clearly shows that the Feng--Kurtz calculation reproduces
the standard MFT Hamiltonian.

\section{Large spin large deviation as MFT on a lattice}
\label{mft}
In this section we explain how the large $s$ large deviation  provides a lattice generalization of various notions in MFT.
We first give a
lattice Martin--Siggia--Rose construction, then derive an explicit rate
function for the density and the current, and finally show that its
continuous-space limit is the usual Gaussian action of MFT.

\subsection{Lattice MSR formulation}
\label{section:msr}
The reduction of the large-$s$ LDP to the MFT can be explicitly understood at the level of deriving actions from a microscopic path integral. One standard way to arrive at the MFT action is to approximate the original microscopic model by a Langevin-like equation and apply the Martin-Siggia-Rose-Janssen-de Dominicis (MSRJD) path integral formalism\cite{MarSigRos73,Jan76,DeDom76}. We can perform a parallel, mathematically regularized calculation directly on the lattice. 

Let us recall the scaled harmonic process $\eta^{(s)}(t) = \frac {1}{2s}\eta(2s t)$ defined in \eqref{scaled-process}. Its microscopic dynamics is governed by the conservation law:

\begin{equation}
\dot\eta^{(s)}_i=j^{(s)}_{i-1,i}-j^{(s)}_{i,i+1},
 \qquad
 j^{(s)}_{i,i+1}(t)
 =\sum_{k\geq1}\frac{k}{2s}
 \bigl(\dot N^+_{i,k}(t)-\dot N^-_{i,k}(t)\bigr),
 \label{sde-harmonic}
\end{equation}
where $N_{i,k}^+$ and $N_{i,k}^-$ denote the number of forward and backward jumps of $k$ particle across the bond $(i, i+1)$ per unit time, respectively. 
Conditionally on the present configuration, these counting processes are independent Poisson  processes with intensities:
\begin{align}
\mathbb{E}[\dot{N}_{i,k}^+(t)\,|\,\eta^{(s)}(t)] &= 
2s\varphi_s(k,2s \eta_i^{(s)}(t)), \qquad\qquad 1\le k \le 2s\eta_i^{(s)}(t),\\
\mathbb{E}[\dot{N}_{i,k}^-(t)\,|\,\eta^{(s)}(t)] &= 2s\varphi_s(k,2s \eta_{i+1}^{(s)}(t)),
\qquad\quad\;\;\, 1\le k \le 2s\eta^{(s)}_{i+1}(t).
\end{align}

Consider the scaled integrated current $Q_T$ 
in the time window $[0,T]$.
To evaluate its generating function, we may introduce 
a Lagrange multiplier $p(t)$ which ensures the conservation law for a path $q(t)$. The path integral representation reads:
\begin{equation}
\langle e^{2s \lambda Q_T} \rangle = \int \mathcal{D}[q] \mathcal D[p]e^{2s \lambda Q_T}  \langle e^{-2s \int_0^T dt \sum_i p_i \left[ \dot{q}_i + \sum_k \frac{k}{2s}\nabla_i(\dot{N}_{i,k}^+ - \dot{N}_{i,k}^-) \right]}\rangle
\end{equation}
Using the summation by parts formula
\begin{equation}
\sum_i f_i (g_{i+1} - g_i) = - \sum_i g_{i+1} (f_{i+1} - f_i) + \text{boundary terms}
\end{equation}
we can shift the spatial difference operator from the current to the conjugate fields. This yields:
\begin{equation}
\langle e^{2s \lambda Q_T} \rangle = \int \mathcal{D}[q] \mathcal{D}[p] e^{2s \lambda Q_T} e^{-2s \int_0^T dt \sum_i p_i \dot{q}_i} \langle e^{\int_0^T dt \sum_i (p_{i+1} - p_i)\sum_{k} {k}(\dot{N}_{i,k}^+ - \dot{N}_{i,k}^-)}\rangle
\end{equation}
For a Poisson process \(N\) with time-dependent intensity \(r(t)\),
\begin{equation}
 \langle\exp\left\{\int_0^T\theta(t)\,dN(t)\right\}\rangle
 =
 \exp\left\{\int_0^T r(t)\bigl(e^{\theta(t)}-1\bigr)\,dt\right\}.
 \label{eq:poisson-exponential-formula}
\end{equation}
Applying this identity, the path integral transforms into:
\begin{eqnarray}
&&\langle e^{2s \lambda Q_T} \rangle \nonumber\\
&&= \int \mathcal{D}[q] \mathcal{D}[p] e^{2s \lambda Q_T} e^{-2s \int_0^T dt \sum_i p_i \dot{q}_i} e^{2s \int_0^T dt \sum_i \left[ \sum_k  \varphi_s(k,2sq_i)(e^{k(p_{i+1}-p_i)} - 1) + \sum_k \varphi_s(k,2sq_{i+1})(e^{-k(p_{i+1}-p_i)} - 1) \right]}\nonumber\\
\end{eqnarray}
This framework provides the lattice MFT Hamiltonian for the harmonic process. 
Performing the exact algebraic summation over $k$ in the limit $s\to\infty$, the Hamiltonian density emerges in a closed logarithmic form:
\begin{equation}
\mathcal{H}(q,p) = - \sum_{i=1}^N \log \left[ 1 + q_i(1 + q_{i+1}) \left( 1 - e^{p_{i+1}-p_i} \right) + q_{i+1}(1 + q_i) \left( 1 - e^{p_i-p_{i+1}} \right) \right]
\end{equation}
which identically matches our main result obtained via the Feng-Kurz generator approach in Eq. \eqref{HamiHarm}.


In the framework of standard MFT, one often writes down the formal
fluctuating hydrodynamic SPDE \eqref{eq:mft-spde}, with $D(\rho)$ and $\sigma(\rho)$ computed from the microscopic model under consideration, 
and applies the MSR scheme to get a path integral representation.  
An outstanding advantage of our lattice-based path integral compared to the standard continuum MFT is that the calculation is performed directly on the original microscopic model, meaning that the stochastic differential equations and the delta-correlated noise are naturally regularized by the lattice spacing. 
This regularized scheme may provide new insights into proving the hydrodynamic large deviations for singular models, such as the symmetric inclusion process, where the continuous density is generally unbounded.

\subsection{Density-current MFT action on a lattice}
\label{subsec:qjaction}

A standard starting point of the conventional MFT is to give an action for the density and the current
\cite{BerDeSGabJonLan15}. 
For a density path $\rho(t,x)$ and current path $j(t,x)$, the dynamical LDP of the empirical density and current $(\rho^{(N)},j^{(N)})$ has the form
\be
\mathbb{P}((\rho^{(N)},j^{(N)})\approx (\rho,j)) \asymp \exp\{-N I(\rho,j)\}
\ee
where the density-current rate functional is
\be
\label{mft-action}
I(\rho,j)=I_0(\rho(0,\cdot))+\frac12 \int_0^T dt \int_0^1 dx \,
\frac{(j+D(\rho)\partial_x \rho)^2}{\sigma(\rho)}
\ee
with the constraint $
\partial_t \rho +\partial_x j=0$. 
If the continuity equation is not satisfied, then $I(\rho,j)=+\infty$.
In the above formulas, $\mathbb{P}$ is the path-space law of the process initialized from its stationary measure, and $I_0$ is the cost of the initial profile.

In this subsection, we derive the corresponding density-current action for the large-\(s\) large deviation on a lattice. 
Let \(q_i(t)>0\) and let \(j_{i,i+1}(t)\in\mathbb R\) be the current across
the oriented bond \((i,i+1)\).  
We have 
\begin{equation}
 \mathbb P_s\bigl((\eta^{(s)},j^{(s)})\simeq(q,j)\bigr)
 \asymp
 \exp\{-2s I(q,j)\}.
 \label{eq:discrete-density-current-LDP}
\end{equation}
where the joint density--current
rate functional is
\begin{equation}
 I(q,j)
 =I_0(q(0))
+\int_0^T\!dt\sum_{i\in\mathbb T_N}
 I_{i,i+1},
 \label{mft-discrete}
\end{equation}
provided
\begin{equation}
 \dot q_i=j_{i-1,i}-j_{i,i+1}.
 \label{eq:discrete-continuity-constraint}
\end{equation}
If the continuity equation is not satisfied, then
\(I(q,j)=+\infty\).  
For the harmonic process, the rate explicitly reads:
\begin{align}
 I_{i,i+1}
 &=j_{i,i+1}\log\!\left[
 \frac{
 j_{i,i+1}(1+q_i+q_{i+1}+2q_iq_{i+1})
 +\sqrt{
 4q_iq_{i+1}(1+q_i)(1+q_{i+1})
 +j_{i,i+1}^{,2}(1+q_i+q_{i+1})^2
 }}
 {2q_i(1+q_{i+1})(1+j_{i,i+1})}
 \right]
 \nonumber\\
 &\quad-
 \log\!\left[
 \frac{
 1+q_i+q_{i+1}+2q_iq_{i+1}
 +\sqrt{
 4q_iq_{i+1}(1+q_i)(1+q_{i+1})
 +j_{i,i+1}^{,2}(1+q_i+q_{i+1})^2
 }}
 {(1+q_i+q_{i+1})^2}
 \right].
\label{eq:explicit-bond-rate}
\end{align}
The expression is understood by continuity at the removable singularity $j_{i,i+1}=-1$.
The bond rate is nonnegative and vanishes at the typical current
\(j_{i,i+1}=q_i-q_{i+1}\).

We derive this formula directly from the relative entropy of the controlled
process.  Under a controlled law, let jumps of size \(k\) from \(i\) to
\(i+1\) and from \(i+1\) to \(i\) have intensities \(2s r^+_{i,k}\) and
\(2s r^-_{i,k}\), respectively.  The relative entropy rate of a Poisson
process with intensity \(2sr\) with respect to one with intensity \(2sc\),
divided by \(2s\), is
\begin{equation}
 h(r\mid c)=r\log\frac{r}{c}-r+c.
 \label{eq:poisson-relative-entropy}
\end{equation}
Therefore
\be 
I_{i,i+1}
=\inf_{\substack{r_{i,k}^\pm\geq0\\
 \sum_{k\geq1}k(r^+_{i,k}-r^-_{i,k})=j_{i,i+1}}}
 \sum_{k\geq1}\Bigl[
 h\bigl(r^+_{i,k}\mid c_k(q_i)\bigr)
 +h\bigl(r^-_{i,k}\mid c_k(q_{i+1})\bigr)
 \Bigr]
 \label{eq:relative-entropy-contraction}
\ee
where
\be
c_k(q_i) = \frac{1}{k} \left(\frac{q_i}{1+q_i}\right)^k, \qquad c_k(q_{i+1}) = \frac{1}{k} \left(\frac{q_{i+1}}{1+q_{i+1}}\right)^k.
\ee
Introducing a multiplier \(\lambda_{i,i+1}\) on each bond, the stationarity condition of the optimization problem gives the optimizer
\begin{equation}
\bar{r}^{+}_{i,k}=c_k(q_i)e^{k\lambda_{i,i+1}},
 \qquad
 \bar{r}^-_{i,k}=c_k(q_{i+1})e^{-k\lambda_{i,i+1}}
 \label{eq:optimal-controlled-rates}
\end{equation}
while the current constraint becomes
\begin{align}
 j_{i,i+1}
 &=
 \frac{q_i e^{\lambda_{i,i+1}}}
 {1+q_i-q_i e^{\lambda_{i,i+1}}}
 -
 \frac{q_{i+1}e^{-\lambda_{i,i+1}}}
 {1+q_{i+1}-q_{i+1}e^{-\lambda_{i,i+1}}}.
 \label{eq:current-multiplier-equation}
\end{align}
Evaluating the relative entropy at the minimizing rates and summing the geometric series gives
\begin{align}
 I_{i,i+1}
 &=
 j_{i,i+1}\lambda_{i,i+1}
 +\log[1-q_i(e^{\lambda_{i,i+1}}-1)]
 +\log[1-q_{i+1}(e^{-\lambda_{i,i+1}}-1)].
 \label{eq:entropy-at-minimizer}
\end{align}
After bringing the two fractions in
\eqref{eq:current-multiplier-equation} to a common denominator, one obtains
the quadratic equation
\begin{align}
 &q_i(1+q_{i+1})(1+j_{i,i+1})e^{2\lambda_{i,i+1}}
 -j_{i,i+1}(1+q_i+q_{i+1}+2q_iq_{i+1})
 e^{\lambda_{i,i+1}}
 +q_{i+1}(1+q_i)(j_{i,i+1}-1)=0.
 \label{eq:quadratic-optimal-multiplier}
\end{align}
The root lying in the convergence interval is
\begin{align}
 e^{\lambda_{i,i+1}}
 ={}&
 \frac{
 j_{i,i+1}(1+q_i+q_{i+1}+2q_iq_{i+1})
 +\sqrt{
 4q_iq_{i+1}(1+q_i)(1+q_{i+1})+j_{i,i+1}^{,2}(1+q_i+q_{i+1})^2
 }}
 {2q_i(1+q_{i+1})(1+j_{i,i+1})}.
 \label{eq:optimal-multiplier}
\end{align}
Substitution of
\eqref{eq:optimal-multiplier} 
into  \eqref{eq:entropy-at-minimizer}
yields \eqref{eq:explicit-bond-rate}.


\subsection{Continuous-space limit of the action}
\label{sec:continuum-limit-action}

We  recover the Gaussian MFT action, in direct analogy with the
Hamiltonian calculation of section \ref{subsec:continuum-hamiltonians}. Expanding the explicit bond rate
around its minimum \(j_{i,i+1}=q_i-q_{i+1}\) gives
\begin{align}
 I_{i,i+1}
 &=
 \frac{
 [j_{i,i+1}-(q_i-q_{i+1})]^2
 }{
 2[q_i(1+q_i)+q_{i+1}(1+q_{i+1})]
 }
+O\!\left(
 |j_{i,i+1}-(q_i-q_{i+1})|^3
 \right).
 \label{eq:bond-rate-quadratic-expansion}
\end{align}
Set \(q_i(t)=\rho(i/N,t)\) and  $j_{i,i+1}(t)=\frac{1}{N} j(i/N,t)\).   For smooth
fields,
\begin{align}
 q_i-q_{i+1}
 &=-\frac{1}{N}\partial_u\rho(u,t)+O(N^{-2}),
 \nonumber\\
q_i(1+q_i)+q_{i+1}(1+q_{i+1})
 &=2\rho(u,t)(1+\rho(u,t))+O(N^{-1}).
\label{eq:continuum-expansions-action}
\end{align}
Therefore, accelerating time by \(N^2\),
\eqref{eq:bond-rate-quadratic-expansion} gives directly
\begin{equation}
 N^2 I_{i,i+1}
 =
 \frac{[j(u,t)+\partial_u\rho(u,t)]^2}
 {4\rho(u,t)(1+\rho(u,t))}
 +O(N^{-1}).
 \label{eq:rescaled-bond-cost-limit}
\end{equation}
Taking the Riemann-sum limit and integrating in time gives
\begin{equation}
 I_{\mathrm{MFT}}(\rho,j)
 =
 \frac{1}{2}\int_0^T\!dt\int_0^1\!du\,
 \frac{
 [j(u,t)+D(\rho)\partial_u\rho(u,t)]^2
 }{
 \sigma(\rho(u,t))
 },
 \label{mft-action2}
\end{equation}
where \(D(\rho)=1\) and \(\sigma(\rho)=2\rho(1+\rho)\); the cost is
\(+\infty\) unless \(\partial_t\rho+\partial_u j=0\).
Thus the standard Gaussian MFT action is the continuous-space limit of the
explicit lattice density--current rate.  In the joint large-spin and
hydrodynamic scaling, the corresponding exponential speed is \(2sN\).

\section{Discrete MFT fluctuation theorem and Einstein relation}
\label{sec:lattice-ft-einstein}

In this section, we investigate the fluctuation theorem
within our large-spin framework of lattice large deviations
and clarify its relation to the conventional continuum MFT.
We show that the fluctuation theorem can be formulated
directly in terms of the lattice Hamiltonian.
We also derive a relation between the average current,
its fluctuations, and the chemical-potential difference,
which reduces to the usual Einstein relation in the
continuous space limit.


\subsection{MFT fluctuation theorem}
We first recall the fluctuation theorem in the context of MFT with Hamiltonian
\begin{equation}
    \mathcal H_{\mathrm{MFT}}(\rho,H)
    =
    \int_{\mathbb T}
    \left\{
    -D(\rho)\,\partial_u\rho\,\partial_uH
    +
    \frac{1}{2}\sigma(\rho)(\partial_uH)^2
    \right\}\,du .
    \label{eq:mft-hamiltonian-ft-einstein}
\end{equation}
Assuming that the equilibrium free energy is local
$    \mathcal F(\rho)
    =
    \int_{\mathbb T} f(\rho(u))\,du 
    $
with chemical potential 
\begin{equation}
    \mu(u)
    =
    \frac{\delta\mathcal F}{\delta\rho(u)}
    =
    f'(\rho(u)),
\end{equation}
the  fluctuation theorem, in Hamiltonian form, is the symmetry
\begin{equation}
    \mathcal H_{\mathrm{MFT}}(\rho,H)
    =
    \mathcal H_{\mathrm{MFT}}(\rho,\mu-H).
    \label{eq:continuum-ft-hamiltonian-new}
\end{equation}
Comparing the coefficients of \(\partial_uH\) and of constant terms in the two sides of
\eqref{eq:continuum-ft-hamiltonian-new} and 
using
$\partial_u\mu=f''(\rho)\partial_u\rho$,
one obtains
\begin{equation}
    D(\rho)
    =
    \frac{1}{2}\sigma(\rho)f''(\rho).
    \label{eq:continuum-einstein-from-ft-new}
\end{equation}
Thus, in the quadratic continuum theory, the fluctuation theorem and the Einstein relation are equivalent.

The fluctuation theorem  can also be expressed in terms of the  density-current cost function \eqref{mft-action2}.
Using the Einstein relation, one finds
\begin{equation}
    I_{\mathrm{MFT}}(\rho,j)
    -
    I_{\mathrm{MFT}}(\rho,-j)
    =
    \int_0^T dt\int_0^1 j\,\partial_u\mu\,du .
    \label{eq:continuum-local-db-new}
\end{equation}
The right hand side is the continuum analogue of local detailed balance: the antisymmetric part of the
current cost under \(j\mapsto -j\) is the current multiplied by the thermodynamic force. Since
\begin{equation}
    \frac{d}{dt}\mathcal F(\rho(t))
    =
    \int_{\mathbb T} j\,\partial_u\mu\,du,
\end{equation}
where the continuity equation and periodic boundary conditions have been used, \eqref{eq:continuum-local-db-new}
integrates to the path-space fluctuation relation
\begin{equation}
    I_{\mathrm{MFT}}(\rho,j)
    -
    I_{\mathrm{MFT}}(\rho^R,j^R)
    =
    \mathcal F(\rho(T))-\mathcal F(\rho(0)).
    \label{eq:continuum-path-ft-new}
\end{equation}
Here
\begin{equation}
    \rho^R(t,u)=\rho(T-t,u),
    \qquad
    j^R(t,u)=-j(T-t,u)
\end{equation}
are the time reversed trajectories.

\subsection{Discrete MFT fluctuation theorem}
Let us now discuss the fluctuation theorem for our discrete MFT. 
For the case of the Harmonic process whose large $s$ Hamiltonian ${\cal H}(q,p)$ is given in \eqref{HamiHarm}, it reads
\begin{equation}
     {\cal H}(q,p)
    =
     {\cal H}(q,\mu-p).
    \label{eq:global-lattice-ft-new}
\end{equation}
This is clearly a discrete analog of \eqref{eq:continuum-ft-hamiltonian-new} in the continuum. Below we see that this is a direct consequence of the local detailed balance. 


Recall that the Hamiltonian is in the form, 
\begin{equation}
    {\cal H}(q,p)
    =
    \sum_{i\in\mathbb T_N}
     {\cal H}_{i,i+1}(q,p_{i+1}-p_i),
    \label{eq:lattice-hamiltonian-ft-new}
\end{equation}
where, denoting $a=p_{i+1}-p_i$, the bond Hamiltonian  
emerged as the sum
\begin{equation}
     {\cal H}_{i,i+1}(q,a)
    =
    \sum_{k\geq 1} c_k(q_i)(e^{ka}-1)
    +
    \sum_{k\geq 1} c_k(q_{i+1})(e^{-ka}-1)
    \label{eq:poisson-bond-hamiltonian-ft-new}
\end{equation}
with $c_k(q_i)
    =
    \frac{1}{k}
    \left(\frac{q_i}{1+q_i}\right)^k$.
Since the reversible measure of the harmonic process is a product of negative binomials, the equilibrium free energy in the variable \(x\) is local and is given by
\begin{equation}
     {\cal F}(q)
    =
    \sum_{i\in\mathbb T_N} f(q_i),
    \qquad
    f(r)=r\log r-(1+r)\log(1+r).
    \label{eq:lattice-free-energy-ft-new}
\end{equation}
Thus the chemical potential at the site $i$ is
\begin{equation}
    \mu_i
    =
    \frac{\partial  {\cal F}}{\partial q_i}
    =
    f'(q_i)
    =
    \log\frac{q_i}{1+q_i}.
    \label{eq:lattice-chemical-potential-ft-new}
\end{equation}

The ratio of the forward and backward \(k\)-jump rates across the bond \((i,i+1)\) is written in terms of this chemical potential as  
\begin{equation}
    \frac{c_k(q_i)}{c_k(q_{i+1})}
    =
    e^{k(\mu_i-\mu_{i+1})},
    \label{eq:lattice-local-detailed-balance-ft-new}
\end{equation}
which is nothing but the local
detailed-balance \cite{khodabandehlou2024local}. 
Writing $b_i:=\mu_{i+1}-\mu_i$,
it is easy to get 
\begin{equation}
     {\cal H}_{i,i+1}(q,a)
    =
     {\cal H}_{i,i+1}(q,b_i-a).
    \label{eq:bond-ft-new}
\end{equation}
Summing over bonds gives the  discrete fluctuation theorem \eqref{eq:global-lattice-ft-new}. 
In terms of the density-current cost, local detailed balance \eqref{eq:lattice-local-detailed-balance-ft-new} is written as
\be
\label{pre-gc}
I(q,j)-I(q,-j)
=
\int_0^T dt \sum_i j_{i,i+1}(\mu_{i+1}-\mu_i),
\ee
which reduces to  
\eqref{eq:continuum-local-db-new} in the continuum limit. 
The Hamiltonian symmetry has the corresponding path-space consequence. Let
\begin{equation}
     {\cal L}(q,\dot q)
    =
    \sup_h
    \left\{
    \sum_i h_i\dot q_i
    -
     {\cal H}(q,p)
    \right\}.
\end{equation}
Using \eqref{eq:global-lattice-ft-new}, with the change of variable \(g=\mu-p\), gives
\begin{equation}
     {\cal L}(q,\dot q)
    -
     {\cal L}(q,-\dot q)
    =
    \sum_i\mu_i\dot q_i
    =
    \frac{d}{dt} {\cal F}(q(t)).
    \label{eq:lattice-lagrangian-ft-new}
\end{equation}
Hence, for the time-reversed path \(q^R(t)=q(T-t)\),
\begin{equation}
    I(q)
    -
    I(q^R)
    =
     {\cal F}(q(T))- {\cal F}(q(0)).
    \label{eq:lattice-path-ft-new}
\end{equation}
Including the equilibrium cost of the initial condition, the total equilibrium path-space rate function is
therefore invariant under time reversal. This is the discrete large-deviation formulation of reversibility.
\begin{remark}[Relation with Gallavotti-Cohen fluctuation theorem]
If \(b_i=b\) is a fixed thermodynamic force and the time-averaged
current satisfies
\[
 \mathbb P\left(\frac{Q_T}{2sT}\simeq j\right)
 \asymp
 \exp\{-2sT\Phi_b(j)\},
\]
then contraction of \eqref{pre-gc} gives the Gallavotti--Cohen current symmetry \cite{kurchan1998fluctuation,lebowitz1999gallavotti}
\be
 \Phi_b(j)-\Phi_b(-j)=bj.
\ee
\end{remark}

\subsection{Einstein relation}
We now explain a consequence of the discrete fluctuation theorem, which becomes the Einstein relation in the continuum limit. 
First differentiate the bond fluctuation theorem \eqref{eq:bond-ft-new} with respect to \(a\) and evaluate at
\(a=0\). This gives the exact identity
\begin{equation}
    \partial_a  {\cal H}_{i,i+1}(q,0)
    =
    -\partial_a  {\cal H}_{i,i+1}(q,b_i).
    \label{eq:exact-derivative-ft-new}
\end{equation}
Expanding further with respect to $b_i(=\mu_{i+1}-\mu_i)$ of the right hand side and noting 

\begin{align}
    \partial_a  {\cal H}_{i,i+1}(q,a)\big|_{a=0}
    &=
\sum_{k\ge1}k[c_k(q_i)-c_k(q_{i+1})]
    = J_{i,i+1}(q), 
    \\
    \partial_a^2  {\cal H}_{i,i+1}(q,a)\big|_{a=0}&=
\sum_{k\ge1}k^2[c_k(q_i)+c_k(q_{i+1})]
=:\sigma^{\mathrm{quad}}_{i,i+1}(q),
\end{align}
we obtain
\begin{equation}
    J_{i,i+1}(q)
    =
    -
    \left[
    J_{i,i+1}(q)
    +   \sigma^{\mathrm{quad}}_{i,i+1}(q)b_i
    +
    O(b_i^2)
    \right].
\end{equation}
Thus
\begin{equation}
    J_{i,i+1}(q)
    =
    -\frac{1}{2}
    \sigma^{\mathrm{quad}}_{i,i+1}(q)(\mu_{i+1}-\mu_i)
    +
    O\bigl((\mu_{i+1}-\mu_i)^2\bigr).
    \label{eq:lattice-linear-response-ft-new}
\end{equation}
This is the linear-response consequence of the discrete fluctuation theorem. 
For the harmonic process, $J_{i,i+1}$ and $\sigma^{\mathrm{quad}}_{i,i+1}$ agree with those obtained directly from the Poisson current
equation:
\begin{equation}
    J_{i,i+1}(q)
    =
    q_i-q_{i+1},
    \label{eq:lattice-current-ft-new}
\end{equation}
and
\begin{equation}
    \sigma^{\mathrm{quad}}_{i,i+1}(q)
    =
    q_i(1+q_i)+q_{i+1}(1+q_{i+1}).
    \label{eq:lattice-sigma-quad-second-ft-new}
\end{equation}
Now, taking the smooth-continuum scaling with 
\begin{equation}
    q_i=\rho(u_i),
    \qquad
    q_{i+1}=\rho(u_i+\varepsilon),
    \qquad
    \varepsilon=\frac{1}{N},
\end{equation}
we see that
\begin{equation}
    \mu_{i+1}-\mu_i
    =
    \varepsilon f''(\rho) \partial_u \rho
    +O(\varepsilon^2),
    \qquad
    f''(\rho)=\frac{1}{\rho(1+\rho)},
    \label{eq:chemical-potential-continuum-ft-new}
\end{equation}
\begin{equation}
\sigma^{\mathrm{quad}}_{i,i+1}(x)
    \longrightarrow
    \sigma(\rho)
    =
    2\rho(1+\rho)
    \label{eq:sigma-continuum-ft-new}
\end{equation}
and hence
\begin{equation}
    J_{i,i+1}(q)
    =
    -\frac{\varepsilon}{2}\sigma(\rho)f''(\rho)
    \partial_u \rho 
    +
    O(\varepsilon^2).
\end{equation}
On the other hand, the diffusive current in the continuum scaling has the Fick form
\begin{equation}
    J_{i,i+1}(q)
    =
    - \varepsilon D(\rho)\partial_u \rho
    +O(\varepsilon^2).
\end{equation}
Comparing the two expressions gives
\begin{equation}
    D(\rho)
    =
    \frac{1}{2}\sigma(\rho)f''(\rho).
    \label{eq:einstein-from-lattice-ft-new}
\end{equation}
This is nothing but the Einstein relation. 

More generally, the fluctuation theorem also constrains
higher cumulants of the bond current. Indeed, the symmetry
$\mathcal H_{i,i+1}(q,a)=\mathcal H_{i,i+1}(q,b_i-a)$ implies
$\partial_a^{2n+1}\mathcal H_{i,i+1}(q,a)|_{a=b_i/2}=0$
for all $n\geq 0$.

\section{Other models}
\label{other}
To demonstrate the universality and algebraic flexibility of our large-spin framework, this section extends the large-$s$ large deviation paradigm to other prominent interacting particle systems: the partial symmetric exclusion process (partial SEP) and the symmetric inclusion process (SIP), and also two models of stochastic heat conduction, the integrable heat conduction model and the Brownian energy process (BEP).  While our primary model, the harmonic process, is deeply tied to the non-compact version of the Ishimori chain, the exclusion and inclusion models allow us to understand the universal nature of the appearance of classical spin chain models. 
By applying the Feng-Kurz asymptotic procedure to these  models, we show how their resulting lattice-MFT Hamiltonians naturally map onto the standard classical $\mathfrak{su}(2)$ and $\mathfrak{su}(1,1)$ Heisenberg spin chains.

\subsection{Processes with partial symmetric exclusion}
\label{sec-exclusion}
In this subsection, we focus on the partial symmetric exclusion process (partial SEP), where the local particle capacity at each site is bounded by $2s$. We introduce the microscopic dynamics rules of this model by specifying its scaled Markovian generator under the large-$s$ scaling. Given a half-integer $s \in \mathbb{N}/2$, let $\{\eta(t), t \ge 0\}$ be the Markov process taking values in $\{0,1,\dots,2s\}^N$. The partial SEP \cite{schutz1994non} is  defined by the generator
\be
L =  \sum_{i=1}^N L_{i,i+1}
\ee
with
\begin{eqnarray}
L_{i,i+1}f(\eta) 
& = & 
\frac{\eta_i}{2s} (1-\frac{\eta_{i+1}}{2s}) (f(\eta - \delta_i + \delta_{i+1})-f(\eta)) \nonumber \\ 
&+&  
\frac{\eta_{i+1}}{2s}(1-\frac{\eta_i}{2s})(f(\eta + \delta_i -\delta_{i+1})-f(\eta))
\end{eqnarray}
For showing the LLN for the process $\{\eta^{(s)}(t), \,t\ge 0\}$ , the scaled generator $L^{(s)}$ reads
\begin{eqnarray}
\label{exclsuion-scaled-gen}
L^{(s)}f(x) =
&&
2s \sum_{i=1}^N \Big(
x_i(1-x_{i+1})(f(x -\frac{\delta_i}{2s}+\frac{\delta_{i+1}}{2s})-f(x)) \nonumber \\ 
&&
\quad\quad\quad+  
x_{i+1}(1-x_{i})(f(x +\frac{\delta_i}{2s}-\frac{\delta_{i+1}}{2s})-f(x)) 
\Big)
\end{eqnarray}
Assuming  that $f:[0,\infty)^N\to\mathbb{R}$ is smooth and has a Taylor expansion, we have
\be
\label{taylor}
f(x -\frac{\delta_i}{2s}+\frac{\delta_{i+1}}{2s})-f(x) = \sum_{\ell=1}^{\infty}\frac{1}{\ell!}  \left(-\frac{1}{2s}\right)^\ell \left(\frac{\partial }{\partial x_i} - \frac{\partial }{\partial x_{i+1}}\right)^{\ell} f(x)
\ee
It follows that, in the limit of large spin,  only the term $\ell=1$ survives  in the expansion of  $L^{(s)}$
\begin{eqnarray}
\label{gen-s-partial}
L^{(s)}f(x) =
&&
\sum_{i=1}^N \left(
\Big(
 -x_i(1-x_{i+1}) +x_{i+1}(1-x_i)
\Big)
 \Big(\frac{\partial }{\partial x_i} - \frac{\partial }{\partial x_{i+1}} \Big) f(x)
 + o(1)
\right) 
\end{eqnarray}
Thus we get
\be
\label{conclusion}
\lim_{s\to\infty} L^{(s)} =L^X ,
\ee
where $L^X$ was defined in \eqref{Lx}.


Next we consider the large deviation. 
The $\{\eta^{(s)}(t), \,t\ge 0\}$ process with generator \eqref{gen-s-partial} 
has a path-space large deviation function of the form \eqref{ld-fct2} 
with Hamiltonian
\be
\label{ld-ham-exclusion}
 {\cal H}_{\text{pSEP}}(q,p) =  \sum_{i=1}^N \Big(
q_i(1-q_{i+1}) (e^{-(p_{i}-p_{i+1})}  - 1)  +   
(1-q_{i})q_{i+1} (e^{-(p_{i+1} -p_{i}) } - 1) \Big)
\ee

The Hamiltonian appearing in the large deviation function is computed as
\be
{\cal H}(q,p) = \lim_{s\to\infty} {\cal H}^{(2s)}(q,p) 
\ee
with
\be
 {\cal H}^{(2s)}(q,p) = \frac{1}{2s} e^{-2s p q} L^{(s)}  e^{2s p q}
\ee
where $p q = \sum_{j=1}^N p_j q_j$. Inserting \eqref{gen-s-partial}
we have
\begin{eqnarray}
 {\cal H}^{(2s)}(q,p)
 & = & 
\frac{1}{2s} e^{-2s p q} 2s \sum_{i=1}^N \Big(
q_i(1-q_{i+1})(e^{2s p(q -  \frac{1}{2s} \delta_i + \frac{1}{2s} \delta_{i+1} ) } - e^{2s p q})   \nonumber\\
&&  \qquad \qquad \qquad
+   
(1-q_{i})q_{i+1} (e^{2s p(q +  \frac{1}{2s} \delta_i - \frac{1}{2s} \delta_{i+1} ) } - e^{2s p q})  \Big)
\end{eqnarray}
Simplifying, we get
\begin{eqnarray}
 {\cal H}^{(2s)}(q,p)
& = & 
\sum_{i=1}^N \Big(
q_i(1-q_{i+1}) (e^{-(p_{i}-p_{i+1})}  - 1)   +   
(1-q_{i})q_{i+1} (e^{-(p_{i+1} -p_{i}) } - 1)  \Big)
\end{eqnarray}
which does not depend on $s$. As a consequence this is also ${\cal H}(q,p)$.

As in the case of the harmonic model, this Hamiltonian can be rewritten in terms of the classical spins. 
Let us set 
\be
\label{su2-classical}
{\cal S}_i^+ = (1-q_i) e^{p_i}\qquad\qquad
{\cal S}_i^-  = q_i e^{-p_i}\qquad\qquad
{\cal S}_i^z = q_i -\frac{1}{2}
\ee
where  $(q_i,p_i)$ are fundamental canonical variables satisfying
$\{q_i,p_i\} = 1$. One can check that 
the spin satisfies the Poisson brackets
\be
 \{{\cal S}_i^z,\pm {\cal S}_j^{\pm}\} = \pm\delta_{i,j} {\cal S}_i^{\pm} \qquad\qquad \{{\cal S}_i^+,{\cal S}_j^-\} = 2\delta_{i,j}{\cal S}_i^z 
\ee
This is the classical $su(2)$ algebra. 
Introducing spin vector ${\cal S}_i=({\cal S}_i^x,{\cal S}_i^y,{\cal S}_i^z)$, with ${\cal S}_j^{x} = ({\cal S}_j^+ +{\cal S} _j^-)/2$ and ${\cal S}_j^{y} = ({\cal S}_j^+ -{\cal S} _j^-)/2i$, one sees that the spin on each site is constrained on the sphere, i.e.
\be
({\cal S}_i^x)^2 + ({\cal S}_i^y)^2 +({\cal S}_i^z)^2  = \frac14,
\ee and the Hamiltonian can be written as the classical XXX spin chain, 
\be
{\cal H} = \sum_{i=1}^N \Big(2 {\cal S}_i \cdot {\cal S}_{i+1} -\frac{1}{2}\Big)
\ee
Indeed, using \eqref{su2-classical}, the scalar product reads
\begin{eqnarray}
2 {\cal S}_i \cdot {\cal S}_{i+1} -\frac{1}{2}
&= & 
 {\cal S}_i^+{\cal S}_{i+1}^- + {\cal S}_i^-{\cal S}_{i+1}^- + 2{\cal S}_i^z {\cal S}_{i+1}^z  - \frac{1}{2} \nonumber \\
& = & 
q_i(1-q_{i+1}) (e^{-(p_{i}-p_{i+1})}  - 1)  +   
(1-q_{i})q_{i+1} (e^{-(p_{i+1} -p_{i}) } - 1)
\end{eqnarray}
Thus we see that the Hamiltonian of the large deviation function for the partial symmetric exclusion process 
coincides with the classical XXX Hamiltonian.

Again this is consistent with the fact that the generator of the partial exclusion process can be written as the XXX quantum spin chain
\be
H = \frac{1}{(2s)^2}\sum_{i=1}^N \left( S_i^+S_{i+1}^- +  S_i^-S_{i+1}^+ + 2S_i^zS_{i+1}^z - 2s^2\right)
\ee
by considering a $(2s+1)-$dimensional representation of the $\mathfrak{su}(2)$ Lie algebra 
\be
\label{su2-commutators}
[S^z, S^{\pm} ] = \pm S^{\pm}, \qquad  [S^+ , S^- ] = 2S^z.
\ee 
Only for spin $s=1/2$ the quantum spin chain is known to be integrable, and in this case one recovers the symmetric simple exclusion process.

\subsection{Symmetric inclusion processes}
\label{sec-inclusion}

Let $\{\eta(t),\, t\ge 0\}$ be the Markov process taking values in $\{0,1,\ldots\}^{N}$ defined by the generator
\be
L =  \sum_{i=1}^N L_{i,i+1}
\ee
with
\begin{eqnarray}
\label{gen-inclusion}
L_{i,i+1}f(\eta) 
& = & 
\frac{\eta_i}{2s} (1+\frac{\eta_{i+1}}{2s}) (f(\eta - \delta_i + \delta_{i+1})-f(\eta)) \nonumber \\ 
&+&  
\frac{\eta_{i+1}}{2s}(1+\frac{\eta_i}{2s})(f(\eta + \delta_i -\delta_{i+1})-f(\eta))
\end{eqnarray}
This process arises from the XXX spin chain with non-compact spins
\be
H = \frac{1}{(2s)^2}\sum_{i=1}^N \left( S_i^+S_{i+1}^- +  S_i^-S_{i+1}^+ - 2S_i^zS_{i+1}^z +2s^2 \right)
\ee
where $S^{\pm},S^z$ are represented as 
 infinite dimensional matrix of the $\mathfrak{su}(1,1)$ Lie algebra \eqref{su11qu} as in the case of the harmonic process.

First one can check that the LLN is exactly the same as for the harmonic and partial exclusion processes. 
Next, we derive the large deviation Hamiltonian for the symmetric inclusion process with generator \eqref{gen-inclusion}. Following the same Feng-Kurz asymptotic procedure $H(q, p) = \lim_{s \to \infty} \frac{1}{2s} e^{-2spq} L^{(s)} e^{2spq}$, we directly find:
\begin{equation}
\mathcal{H}_{\text{SIP}}(q, p) = \sum_{i=1}^N \left( q_i(1 + q_{i+1})(e^{-(p_i - p_{i+1})} - 1) + (1 + q_i)q_{i+1}(e^{-(p_{i+1} - p_i)} - 1) \right).
\label{ld-ham-inclusion}
\end{equation}
Compared to the partial exclusion case \eqref{ld-ham-exclusion}, the terms $(1 - q_i)$ are naturally replaced by $(1 + q_i)$, reflecting the attractive and unbounded nature of the inclusion dynamics.


Under the $su(1,1)$ spin variables \eqref{su11pq} as for the harmonic process, the inclusion Hamiltonian \eqref{ld-ham-inclusion} is exactly rewritten as:
\begin{equation}
\mathcal{H}_{\text{SIP}} = \sum_{i=1}^N \left( -2 \mathcal{S}_i \cdot \mathcal{S}_{i+1} + \frac{1}{2} \right)
\label{inclusion-XXX-noncompact}
\end{equation}
where the scalar product reads $2\mathcal{S}_i \cdot \mathcal{S}_{i+1} = 2\mathcal{S}^z_i \mathcal{S}^z_{i+1} - \mathcal{S}^+_i \mathcal{S}^-_{i+1} - \mathcal{S}^-_i \mathcal{S}^+_{i+1}$. This shows that the large-spin large deviation of the symmetric inclusion process is governed by the classical non-compact XXX spin chain, establishing a neat symmetry with the exclusion process governed by the standard compact XXX chain.

It is worth emphasizing that in the continuous MFT for the inclusion process, the field equations suffer from severe singularities due to the unbounded nature of the continuous density. In our large-$s$ lattice framework, however, the spatial correlations and local density fluctuations are naturally suppressed and regularized by the rigid lattice spacing. The fact that $\mathcal{H}_{\text{SIP}}$ maps perfectly to the well-behaved classical non-compact XXX spin chain strongly suggests that the large-$s$ LDP on a fixed lattice is mathematically robust, offering a strategic advantage over the standard continuum KOV approach.

\subsection{Integrable heat conduction model}
\label{subsec:integrable_heat_conduction}

As another application of our general framework, we consider the integrable heat conduction model parameterized by $s > 0$, which governs the continuous energy exchange on the one-dimensional torus $\mathbb{T}_N = \mathbb{Z} / N\mathbb{Z}$ \cite{franceschini2023integrable}. 

The state space of the system is given by $\mathcal{X}_N = [0, \infty)^{\mathbb{T}_N}$. A configuration is denoted by $x = (x_i)_{i \in \mathbb{T}_N}$, where $x_i \ge 0$ represents the continuous energy (or mass) at site $i$. The microscopic generator of the bulk dynamics $L$ is given by:
\begin{equation}
    L f(x) = \sum_{i \in \mathbb{T}_N} \left( {L}_{i, i+1}^{\to} + {L}_{i, i+1}^{\leftarrow} \right) f(x),
\end{equation}
where the asymmetric local transfer operators are defined as:
\begin{align}
    {L}_{i, i+1}^{\to} f(x) &= \int_{0}^{x_i} \frac{d\alpha}{\alpha} \left( 1 - \frac{\alpha}{x_i} \right)^{2s-1} \left[ f(x - \alpha \delta_i + \alpha \delta_{i+1}) - f(x) \right], \\
    {L}_{i, i+1}^{\leftarrow} f(x) &= \int_{0}^{x_{i+1}} \frac{d\alpha}{\alpha} \left( 1 - \frac{\alpha}{x_{i+1}} \right)^{2s-1} \left[ f(x + \alpha \delta_i - \alpha \delta_{i+1}) - f(x) \right].
\end{align}

To analyze the classical limit, we consider the large $s$ regime. Under the scaling where the configuration is rescaled as $x \to 2s x$ and  is speeded up by a factor of $2s$, the empirical energy profile $x^{(s)}(t)$ converges in probability as $s \to \infty$ to the deterministic limit. Specifically, the Law of Large Numbers (LLN) is established in this large $s$ limit, showing that the scaled energy profile $x^{(s)}(t) = (x^{(s)}_i(t))_{i \in \mathbb{T}_N}$ converges to the unique solution of the system of coupled ordinary differential equations \eqref{odee}.

To establish the large deviations rate function, we again employ the Feng-Kurz scheme \cite{FenKur06}. 
The result is 
\begin{equation}
\label{IHC_hami_cl}
    {\cal H}(q, p) = - \sum_{i \in \mathbb{T}_N} \left[ \log \Big( 1 - q_i (p_{i+1} - p_i) \Big) + \log \Big( 1 - q_{i+1} (p_i - p_{i+1}) \Big) \right].
\end{equation}

As in the case of the harmonic process, 
the Hamiltonian for the integrable heat conduction model \eqref{IHC_hami_cl} can also be written as the Ishimori spin chain Hamiltonian \eqref{HamiIshi},
if we use a representation of the classical non-compact $\mathfrak{su}(1,1)$ spin variables in terms of $p,q$ variables at each site $i$ as:
\begin{equation}
    \mathcal{S}_i^+ = q_i p_i^2 + p_i, \quad \mathcal{S}_i^- = q_i, \quad \mathcal{S}_i^z = q_ip_i + \frac12.
    \label{su11pq2}
\end{equation}
Note that this is different from \eqref{su11pq} used for mapping the harmonic Hamiltonian to Ishimori one.

Again this is consistent with the fact that the generator of the integrable heat conduction model is written as the non-compact $su(1,1)$ quantum spin chain \eqref{HamiHarm} with the representation 
\begin{equation}
    S^z_i = x_i \frac{\partial}{\partial x_i} + s, \quad S^+_i = x_i, \quad S^-_i = x_i \frac{\partial^2}{\partial x_i^2} + 2s \frac{\partial}{\partial x_i}.
\end{equation}

\subsection{The Brownian Energy Process}
Another model with continuous local state spaces is the Brownian Energy Process (BEP), which is a diffusion process representing the conservative exchange of continuous energy variables on a lattice and is defined as follows \cite{giardina2009duality}. 
Let $\{x(t), t \ge 0\}$ be the Markov process taking values in the continuous state space $[0, \infty)^N$, where $x_i(t)$ represents the energy at site $i$ at time $t$. For a given parameter $s > 0$, the generator of the Brownian energy process $\text{BEP}(2s)$ is given by:
\be
L = \sum_{i=1}^N L_{i,i+1}
\ee
where the bond generator $L_{i,i+1}$ acts on smooth test functions as:
\be
L_{i,i+1} = x_i x_{i+1} \left( \frac{\partial}{\partial x_i} - \frac{\partial}{\partial x_{i+1}} \right)^2 + 2s (x_{i+1} - x_i) \left( \frac{\partial}{\partial x_i} - \frac{\partial}{\partial x_{i+1}} \right).
\ee
By scaling the variables and the generator appropriately under the large-$s$ limit, i.e. $x^{(s)}(t)= \frac{x(2s t)}{2s}$, the stochastic energy profile concentrates in the limit $s\to\infty$ around its deterministic macroscopic trajectory, satisfying the lattice law of large numbers (LLN) governed by the discrete heat equation \eqref{odee}.

For studying large deviation for large $s$, 
we again apply the Feng-Kurtz formula to the scaled generator of the BEP. 
The resulting classical Hamiltonian is 
 \be
 \mathcal{H}_{\text{BEP}}(q,p) = \sum_{i=1}^N \left[ q_i q_{i+1} (p_{i+1} - p_i)^2 - (q_{i+1} - q_i)(p_{i+1} - p_i) \right].
 \ee
We can again express the BEP Hamiltonian in terms of classical spin variables. 
In terms of the classical $su(1,1)$ spin variables
\eqref{su11pq2} the Hamiltonian is written as the classical XXX spin chain \eqref{inclusion-XXX-noncompact} due to 
\be
-2\mathcal{S}_i \cdot \mathcal{S}_{i+1} +\frac12 = -2\mathcal{S}_i^z \mathcal{S}_{i+1}^z + \mathcal{S}_i^+ \mathcal{S}_{i+1}^- + \mathcal{S}_i^- \mathcal{S}_{i+1}^+ +\frac12
=
q_i q_{i+1} (p_{i+1} - p_i)^2 - (q_{i+1} - q_i)(p_{i+1} - p_i).
\ee

\appendix
\renewcommand{\thesection}{Appendix \Alph{section}}
\renewcommand{\theequation}{\Alph{section}.\arabic{equation}}

\section{The example of independent random walkers}
\label{rw}
In this appendix, we show results of application of our arguments to the simplest example of particle systems, i.e. independent random walkers. 
This may be useful for understanding the general scheme. 

Denote by $\{\eta(t), t\ge 0\}$ a system of independent random walks moving on the one-dimensional lattice
and jumping symmetrically to their nearest neighbors at rate 1.
This is a Markov process  taking values in $\{0,1,2,\ldots\}^{N}$ and defined by the generator
\be
L f(\eta) =  \sum_{i=1}^N \Big(\eta_i (f(\eta - \delta_i + \delta_{i+1})-f(\eta)) + \eta_{i+1}(f(\eta + \delta_i -\delta_{i+1})-f(\eta))\Big).
\ee
Here $\eta=(\eta_i)_{1\le i \le N}$ is a configuration with $\eta_i$ particles at site $i$, $\delta_i$ is a configuration
with just one particle at site $i$ and  $f:\mathbb{N}^{N}\to \mathbb{R}$ is a function in the domain of the generator.
Without loss of generality we assume periodic boundary conditions.
Now introduce a scaling parameter $n\in\mathbb{N}$  and consider the sequence of processes $\{\eta^{(n)}(t), \, t\ge 0\}$ 
defined by
\be
\eta^{(n)}_{i}(t) = \frac{\eta_i(t)}{n}. 
\ee
This is a process taking values in $\{0,\frac{1}{n},\frac{2}{n},\ldots\}^{N}$ and defined by the generator
\be
\label{Generator-irw-n}
L^{(n)} f(x) = n \sum_{i=1}^N \left( x_i \Big( (f(x - \frac{1}{n}{\delta_i} + \frac{1}{n}\delta_{i+1})-f(x)\Big) + x_{i+1}\Big(f(x + \frac{1}{n}{\delta_i} - \frac{1}{n}\delta_{i+1})-f(x)\Big) \right).
\ee
In the limit $n\to\infty$, the limiting process will follow with overwhelming probability a deterministic trajectory $\{x(t),\, t\ge 0\}$ which is the solution of the system of ordinary differential equations 
\be
\label{odeeA}
\dot{x}_i(t) = x_{i-1}(t) - 2x_i(t) + x_{i+1}(t),  \qquad\qquad i \in \{1,2,\ldots,N\}.
\ee
This functional version of the law of large number can be obtained by showing that the generator $L^{(n)}$ converges to 
the generator $L^X$ in \eqref{Lx}.
For the independent random walks example one has
\be
\label{Hamiltonian-irw}
{\cal H} (q,p) = \sum_{i=1}^N \Big( q_i (e^{(p_{i+1}-p_i)}-1) + q_{i+1}(e^{-(p_{i+1}-p_i)}-1)\Big).
\ee

While establishing rigorously the large deviation principle in path space for a general interacting particle system is highly demanding, 
for the characterization of the large deviation function there is a well-established procedure, see  \cite{FenKur06, Fle85}.  
The procedure relies on the introduction of the generator ${\cal H}^{(n)}$ of a non-linear semigroup
\be
\label{H_n}
{\cal H}^{(n)} f := \frac{1}{n} e^{-nf} L^{(n)}  e^{nf} 
\ee
and then the Hamiltonian ${\cal H}$ is given by
\be
\label{H}
{\cal H} = \lim_{n\to\infty}{\cal H}^{(n)}.
\ee
The rationale behind this procedure is that, as the semigroup associated to the generator $L^{(n)}$ gives expectation at time $t$
of a function of the Markov process $\eta^{(n)}(t)$
\be
e^{tL^{(n)}}f(x) = \mathbb{E} \big(f(\eta^{(n)}(t))\,|\, \eta^{(n)}(0) = x\big),
\ee
in the same manner the non-linear semigroup associated to the generator ${\cal H}^{(n)}$ gives expectation at time $t$
of an exponential function of the Markov process $x_n(t)$
\be
V^{(n)}(t)f(x) =  \frac{1}{n} \log \Big( \mathbb{E} \big(e^{nf(\eta^{(n)}(t))} \,|\, \eta^{(n)}(0) = x\big)\Big).
\ee
One can check that the application of the general procedure given in \eqref{H_n} and \eqref{H} to the independent random walks generator $L^{(n)}$  in \eqref{Generator-irw-n} immediately yields the Hamiltonian  \eqref{Hamiltonian-irw}.
We also remark that the canonical change of variables
\be
Q_i = q_i e^{-p_i} \qquad\qquad P_i = e^{p_i}
\ee
gives
\be
\label{Hamiltonian-irw-QP}
{\cal H} (Q,P) = - \sum_{i=1}^N \Big( (Q_i-Q_{i+1})(P_i-P_{i+1}) \Big). 
\ee

\bibliographystyle{unsrt}
\bibliography{ref}

\end{document}